\documentstyle[aps,preprint]{revtex}
\begin{document}
\draft

\title{\large $^{8}$Be cluster emission versus $\alpha$ evaporation in
$^{28}$Si + $^{12}$C}

\author{M.~Rousseau\thanks{Corresponding author. Electronic address:
marc.rousseau@ires.in2p3.fr}, C.~Beck, C.~Bhattacharya\thanks{Permanent
address: VECC, 1/AF Bidhan Nagar, Kolkata 64, India.}, V.~Rauch, O.~Dorvaux,
K.~Eddahbi, C.~Enaux, R.M.~Freeman, F.~Haas, D.~Mahboub{\thanks{Present
address: University of Surrey, Guildford GU2 7XH, United Kingdom.}}, R. 
Nouicer\thanks{Present address: Department of Physics, University of Illinois
at Chicago, Chicago, Illinois 60607-7059, USA.}, P.~Papka,
O.~Stezowski{\thanks{Permanent address: IPN Lyon, F-69622 Villeurbanne,
France.}}, and S.~Szilner } 

\address{\it Institut de Recherches Subatomiques, UMR7500, Institut National de
Physique Nucl\'eaire et de Physique des Particules - Centre National de la
Recherche Scientifique/Universit\'e Louis Pasteur, 23 rue du Loess, B.P. 28,
F-67037 Strasbourg Cedex 2, France}

\author{A.~Hachem and E.~Martin}

\address{\it Universit\'e de Nice-Sophia Antipolis, F-06108 Nice, France}

\author{S.J.~Sanders and A.K.~Dummer{\thanks{Present address: Triangle
Universities Nuclear Laboratory, University of North Carolina, Durham, NC
27708-0308, USA}}} 

\address{\it Department of Physics and Astronomy, University of Kansas,
Lawrence, Kansas 66045, USA}

\author{A.~Szanto de Toledo}

\address{\it Departamento de F\'{\i}sica Nuclear, Instituto de F\'{\i}sica da
Universidade de S\~ao Paulo, C.P. 66318-05315-970 - S\~ao Paulo, Brazil} 

\date{\today}
\maketitle

\newpage

\begin{abstract}

{The possible occurence of highly deformed configurations in the $^{40}$Ca
di-nuclear system formed in the $^{28}$Si + $^{12}$C reaction is investigated
by analyzing the spectra of emitted light charged particles. Both inclusive and
exclusive measurements of the heavy fragments (A $\geq$ 10) and their
associated light charged particles (protons and $\alpha$ particles) have been
made at the IReS Strasbourg {\sc VIVITRON} Tandem facility at bombarding
energies of $E_{lab}$($^{28}$Si) = 112 MeV and 180 MeV by using the {\sc ICARE}
charged particle multidetector array. The energy spectra, velocity
distributions, in-plane and out-of-plane angular correlations of light charged
particles are compared to statistical-model calculations using a consistent set
of parameters with spin-dependent level densities. This spin dependence
approach suggests the onset of large nuclear deformation in $^{40}$Ca at high
spin. This conclusion might be connected with the recent observation of
superdeformed bands in the $^{40}$Ca nucleus. The analysis of $\alpha$
particles in coincidence with $^{32}$S fragments suggests a surprisingly strong
$^{8}$Be cluster emission of a binary nature.}

\end{abstract}
{PACS number(s): 25.70.Gh, 25.70.Jj, 25.70.Mn, 24.60.Dr}

\newpage

\section{INTRODUCTION}

The formation and binary decay of light nuclear systems in the A$_{CN}$
$\leq$ 60 mass region produced by low-energy (E$_{lab}$ $\leq$ 7 MeV/nucleon)
heavy-ion reactions has been extensively studied both from the experimental and
the theoretical points of view \cite{Sanders99}. In most of the reactions
studied the binary breakup of the compound nucleus (CN) is seen as either a
fusion-fission (FF) \cite{Sanders99,Matsuse97} or a deep-inelastic (DI)
orbiting \cite{Shivakumar87} process. The large-angles orbiting yields are
found to be particularly strong in the $^{28}$Si + $^{12}$C reaction
\cite{Shapira82}, as illustrated by Fig.~1 which summarizes some of the
experimental results that have been collected for this system; i.e., orbiting
cross sections~\cite{Shapira82,Shapira84} and total evaporation residue (ER)
cross
sections~\cite{Gary82,Lesko82,Nagashima82,Harmon86,Harmon88,Vineyard93,Arena94}.
Since many of the conjectured features for orbiting yields are similar to those
expected for the FF mechanism, it is difficult to fully discount FF as a
possible explanation for the large energy-damped $^{28}$Si + $^{12}$C
yields~\cite{Shivakumar87,Shapira82,Shapira84}. However, FF
calculations~\cite{Sanders99} significantly underpredict the cross sections
measured in the carbon channel by almost a factor of 3, thus suggesting an
alternative mechanism (See Fig.~1). FF, DI orbiting, and even molecular-resonance behavior
may all be active~\cite{Pocanic85} in the large-angle yields of the $^{28}$Si +
$^{12}$C reaction~\cite{Ost79,Barrette79}. The back-angle elastic scatterings of
$^{28}$Si ions from $^{12}$C displays structured excitation functions and
oscillatory angular distributions in agreement with the relatively weak
absorption of this system \cite{Beck94}. Moreover, the resonant gross
structure~\cite{Ost79} is fragmented into very striking intermediate width
resonant structure~\cite{Barrette79}. 

Superdeformed (SD) rotational bands have been found in various mass regions (A
= 60, 80, 130, 150 and 190) and, very recently, SD bands have also been
discovered in the N = Z nuclei $^{36}$Ar~\cite{Svensson00,Svensson01} and
$^{40}$Ca~\cite{Ideguchi01}. These new results make the A$_{CN}$ $\approx$ 40
mass region of particular interest since quasimolecular resonances have also
been observed in both the $^{36}$Ar and $^{40}$Ca dinuclear
systems~\cite{Pocanic85}. Although there is no experimental evidence to link
the SD bands with the higher lying rotational bands formed by known
quasimolecular resonances, both phenomena are believed to originate from highly
deformed configurations of these systems. Since the detection of light charged
particles (LCP) is relatively simple, the analysis of their spectral shapes is
another good tool for exploring nuclear deformation and other properties of hot
rotating nuclei at high angular momenta. Experimental evidence for
angular-momentum-dependent spectral shapes has already been extensively
discussed in the 
literature~\cite{Choudhury84,Majka87,Govil87,Fornal88,Viesti88,Fornal89,Larana89,Huizenga89,Fornal91a,Agnihotri93,Govil98,Bandyopadhyay99,Govil00a,Govil00b,Bandy01}
and, in particular, the $^{24}$Mg + $^{16}$O reaction~\cite{Fornal91b}, which
reaches the $^{40}$Ca CN, has been studied in detail. Strong deformation
effects were also deduced from angular correlation data for the fusion reaction
$^{28}$Si($^{12}$C,2$\alpha$)$^{32}$S$_{g.s.}$ at E$_{lab}$ = 70 
MeV~\cite{Alamanos83}. We decided to investigate the $^{40}$Ca nucleus produced
through the $^{28}$Si + $^{12}$C reaction at beam energies of E$_{lab}$ = 112
MeV and 180 MeV. As can be observed in Fig.~1, at the lowest incident energy
of the present work the orbiting process is dominant whereas at E$_{lab}$ = 180
MeV a large part of the O and N fully-damped yields may also result from a FF
mechanism. It is interesting to note that the lower energy, E$_{c.m.}$ = 33.6
MeV, corresponds to a well known quasimolecular resonance (see Fig.~6 of
Ref.~\cite{Beck94}). In this article we will focus on the LCP's found in
coincidence with heavy fragments. Data have also been collected for the
$^{28}$Si + $^{28}$Si reaction (leading to the N = Z nucleus $^{56}$Ni) in the
same experimental conditions at two bombarding 
energies~\cite{Bhattacharya99,Bhattacharya02,Rousseau00,Beck00,Rousseau01a}. 

The present paper is organized in the following way. Sec. II describes the
experimental procedures and the data analysis. Sec. III presents the inclusive
and the exclusive $^{28}$Si + $^{12}$C data (part of the experimental results
presented here in detail have already been briefly reported elsewhere
~\cite{Rousseau00,Beck00,Rousseau01a,Rousseau01b,Bhattacharya01}). The data are
analyzed using the Hauser-Feshbach evaporation code {\sc
CACARIZO}~\cite{Choudhury84,Majka87,Viesti88} using a consistent set of
parameters which has been found to succesfully reproduce $^{24}$Mg + $^{16}$O
reaction results~\cite{Fornal91b}. The full statistical-model calculations,
using Monte Carlo techniques to account for the experimental acceptance when
comparing to the experimental exclusive data, are presented in Sec. IV. The
unexpected strong cluster emission of $^{8}$Be as due to a binary decay process
is also discussed in this section. We end with a conclusion in Sec. V. 

\newpage

\section{EXPERIMENTAL PROCEDURES AND DATA ANALYSIS}

\subsection{Experimental procedures}

The experiments were performed at the {\sc VIVITRON} Tandem facility of the
IReS Strasbourg laboratory using 112 MeV and 180 MeV $^{28}$Si beams which were
incident on $^{12}$C targets (160 and 180 $\mu$g/cm${^2}$ thick, respectively)
mounted in the ICARE scattering chamber~\cite{Belier94,Bellot97}. The effective
thicknesses of the $^{12}$C targets were accurately determined using Rutherford
back scattering (RBS) techniques with $^{1}$H and $^{4}$He beams provided by
the Strasbourg 4 MV Van de Graaff
accelerator~\cite{Bhattacharya99,Bhattacharya02,Rousseau01a}. The carbon
buildup corrections were found to be less than 2$\%$ of the total of C atoms in
the targets. Both the heavy fragments (A $\geq$ 10) and their associated LCP's
(protons and $\alpha$ particles) were detected in coincidence using the {\sc
ICARE} charged-particle multidetector array \cite{Belier94,Bellot97} which
consists of nearly 40 telescopes. Inclusive data have also been collected for
heavy fragments and LCP's and presented in Sec.~III.A. 

\subsubsection{Experimental setup at E$_{lab}$ = 112 MeV}

For the measurement at E$_{lab}$($^{28}$Si) = 112 MeV, the heavy fragments
consisting of ER as well as quasi-elastic, deep-inelastic, and fusion-fision
fragments, were detected in 8 gas-silicon hybrid telescopes (IC), each
composed of a 4.8 cm thick ionization chamber, with a thin Mylar entrance
window, followed by a 500 $\mu$m thick Si(SB) detector. The IC's were located
at $\Theta_{lab}$ = $\pm$15$^\circ$, -20$^\circ$, $\pm$25$^\circ$, -30$^\circ$,
-35$^\circ$, and -40$^\circ$ in two distinct reaction planes (for each plane,
the positive and negative angles are defined in a consistent manner as for the
LCP detectors described below). The in-plane detection of coincident LCP's was
done using 4 three-element telescopes (TL3) (40 $\mu$m Si, 300 $\mu$m Si, and 2
cm CsI(Tl)) placed at forward angles ($\Theta_{lab}$ = +15$^\circ$, +25
$^\circ$, +35$^\circ$, and +45$^\circ$), 16 two-element telescopes (TL2) (40
$\mu$m Si, 2 cm CsI(Tl)) placed at forward and backward angles (+40$^\circ$
$\leq$ $\Theta_{lab}$ $\leq$ +115$^\circ$) and, finally, two other IC
telescopes located at the most backward angles $\Theta_{lab}$ = +130$^\circ$
and +150$^\circ$. The CsI(Tl) scintillators were coupled to photodiode
readouts. The IC's were filled with isobutane at a pressure of 30 Torr for the
backward angle telescopes and of 60 Torr for the forward angle detectors, thus
allowing for the simultaneous measurement of both light and heavy fragments. 

\subsubsection{Experimental setup at E$_{lab}$ = 180 MeV}

For the measurement at E$_{lab}$($^{28}$Si) = 180 MeV, three distinct reaction
planes were defined. Two for in-plane correlations and a third one,
perpendicular to the LCP detection plane, for out-of-plane correlation
measurements. The heavy fragments were detected in 10 IC's located at
$\Theta_{lab}$ = $\pm$10$^\circ$, $\pm$15$^\circ$, $\pm$20$^\circ$, and
$\pm$25$^\circ$ ($\Phi_{lab}$ = 0$^\circ$) for the in-plane coincidences, and
at $\Theta_{lab}$ = +10$^\circ$ and +20$^\circ$ ($\Phi_{lab}$ = 90$^\circ$) for
the out-of-plane coincidences. Both the in-plane and out-of-plane coincident
LCP's were detected using 3 TL3's placed at forward angles (2 at $\Theta_{lab}$
= +30$^\circ$ and one at $\Theta_{lab}$ = +35$^\circ$) and 24 TL2's placed at
forward and backward angles (+40$^\circ$ $\leq$ $\Theta_{lab}$ $\leq$
+95$^\circ$). The IC's were filled with isobutane at a pressure of 60 Torr. 

The acceptance of each telescope was defined by thick aluminium collimators.
The distances of these telescopes from the target ranged from 10.0 to 30.0 cm,
and the solid angles varied from 1.0 msr at the most forward angles to 5.0 msr
at the backward angles, according to the expected counting rates. 

\subsection{Data analysis}

The energy calibrations of the different telescopes of the {\sc ICARE}
multidetector array were done using radioactive $^{228}$Th and $^{241}$Am
$\alpha$-particle sources in the 5-9 MeV energy range, a precision pulser, and
elastic scatterings of 112 MeV and 180 MeV $^{28}$Si from $^{197}$Au,
$^{28}$Si, and $^{12}$C targets in a standard manner. In addition, the
$^{12}$C($^{16}$O,$\alpha$)$^{24}$Mg$^{*}$ reaction at E$_{lab}$ = 53 MeV
\cite{Rousseau01a} was used to provide known energies of $\alpha$ particles 
feeding the $^{24}$Mg excited states, thus allowing for calibration of the
backward angle detectors. The proton calibration was achieved using scattered
protons from Formvar targets bombarded in reverse kinematics reactions with both
$^{28}$Si and $^{16}$O beams. On an event-by-event basis, corrections were
applied for energy loss of heavy fragments (A $\geq$ 10) in the targets and in
the entrance window Mylar foils of the IC's and thin Al-Mylar foils of the Si
diodes, and for the pulse height defect in the Si detectors. The IC energy
thresholds and energy resolution for heavy fragments are better than 1.5
MeV/nucleon and 0.7$\%$, respectively, as shown in Fig.~2 for the C and O exit
fragments for the $^{28}$Si(112 MeV) + $^{12}$C reaction at $\Theta_{lab}$ =
15$^{\circ}$. The total energy resolution of 8.78 MeV $\alpha$ particles from
thorium sources has been found to be better than 2.2$\%$ for both the
three-element and two-element light-ion CsI(Tl) telescopes. Absolute cross
sections of inclusive measurements could be obtained within 10-12$\%$ error
bars as due to 3-5$\%$ uncertainties in the target thickness and to 8-10$\%$ 
uncertainties in the electronic deadtime corrections. More details on the
experimental setup of {\sc ICARE} and on the analysis procedures can be found
in Refs.~\cite{Bhattacharya99,Bhattacharya02,Beck00,Rousseau01a,Rousseau01b}
and references therein. 

\newpage

\section{EXPERIMENTAL RESULTS}

\subsection{Inclusive data}

\subsubsection{Heavy fragments}

The fragments with Z = 15-17 have typical inclusive energy spectra of ER's
which are not displayed here, but their LCP exclusive data are discussed in
Sec. III.B. We rather focuss on inclusive data of the binary fragments with Z =
5-14. The C and O fragment energy spectra measured at $\Theta_{lab}$ =
15$^{\circ}$ for the $^{28}$Si + $^{12}$C reaction at E$_{lab}$ = 112 MeV are
displayed in Fig.~2. As expected they are very similar to the spectra measured
by Shapira {\it et al.}~\cite{Shapira82} for the same reaction at E$_{lab}$ =
115 MeV. The peak identifications are given for the ground states as well as
for the single and mutual excited states for the inelastic and
$\alpha$-transfer exit-channels, respectively. The measured yields for
different binary-channel products (Z = 5 to 11) were converted to
center-of-mass cross sections by integrating their energy spectra. The
resulting cross sections are shown in Fig.~3 and Fig.~4 for the two bombarding
energies E$_{lab}$ = 112 MeV and 180 MeV, respectively. In Fig.~3 the present
data (open symbols) are in fairly good agreement with the previously measured
C, N, and O angular distributions of Shapira {\it et al.}~\cite{Shapira82}
(full symbols) obtained at E$_{lab}$ = 115 MeV. All angular distributions of
Figs.~3 and 4 vary as 1/sin$\theta_{c.m.}$, providing strong evidence that the
damped reaction component (of a fusion-fission or DI orbiting origin) has a
long lifetime, thus for these light fragment (Z = 5 to 11) a fast DI mechanism
seems unlikely to contribute significantly. On the other hand such a fast DI
mechanism does appear to be significant for the heavier fragments (Z = 12 to
14). 

By integrating their 1/sin$\theta_{c.m.}$ angular distributions, fully-damped
cross sections for the binary-channel products can be obtained. The results of
the experimental cross sections are given in Table \ref{table1} and compared
with the previously measured excitation functions~\cite{Shapira82,Shapira84} in
Fig.~1. The agreement between both sets of data is good. However, at the
highest bombarding energy E$_{lab}$ = 180 MeV (E$_{c.m.}$ = 54 MeV) a rather
small systematic disagreement can be observed. This 10$\%$ discrepancy is
attributed to uncertainties in the target thickness and electronic deadtime
corrections. 

\subsubsection{Light charged particles}

Typical inclusive energy spectra of $\alpha$ particles are shown in Fig.~5 at
the indicated angles for the $^{28}$Si(180 MeV) + $^{12}$C reaction. The solid
points (with error bars visible when greater than the size of the points) are
the experimental data whereas the solid and dashed lines are statistical model
calculations discussed in Sec.~IV.B. The $\alpha$-particle energy spectra have
a Maxwellian shape with an exponential fall-off at high energy which reflects a
relatively high effective temperature (T$_{slope}$ $\approx$ [8E$^{*}_{\small
CN}$/A$_{\small CN}$]$^{1/2}$ = 3.67 MeV) of the decaying nucleus. The spectral
shape and high-energy slopes are also found to be essentially independent of
angle in the c.m. system. These observations suggest a statistical deexcitation
process arising from a thermalized source such as the $^{40}$Ca CN. Similar
$\alpha$-particle spectra have been measured for the lowest bombarding energy.
These are also consistent with a previous study at E$_{lab}$ = 150
MeV~\cite{Kildir95}. The velocity spectra of protons and $\alpha$ particles,
presented next, are strongly supportive of a statistical deexcitation process. 

The velocity contour maps of the LCP Galilean-invariant differential cross
sections (d$^{2}\sigma$/d$\Omega$dE)p$^{-1}$c$^{-1}$ as a function of the LCP
velocity are known to provide an overall picture of the reaction pattern. From
this pattern the velocity of the emission source can be determined in order to
better characterize the nature of the reaction mechanism. Fig.~6 shows such
typical plots of invariant cross-section in the (V$_{\parallel}$,V$_{\perp}$)
plane for $\alpha$ particles (left side) and protons (right side),
respectively, measured in singles mode. The symbols V$_{\parallel}$ and
V$_{\perp}$ denote laboratory velocity components parallel and perpendicular to
the beam, respectively. For the sake of clarity the velocity cutoffs arising
from the detector low-energy thresholds are indicated for each of the
telescopes. The present data at E$_{lab}$ = 112 and 180 MeV have the same
trends as the plots measured at E$_{lab}$ = 150 MeV (see Fig.~1 of
Ref.~\cite{Kildir95}). The dashed circular arcs, centered on the center-of-mass
V$_{c.m.}$ and defined to visualize the maxima of particle velocity spectra,
describe the data trends rather well. They have radii very close to the Coulomb
velocities of $\alpha$ particles and protons in the decay of $^{40}$Ca$^{*}$
$\rightarrow$ $^{36}$Ar + $^{4}$He, and of $^{40}$Ca$^{*}$ $\rightarrow$
$^{39}$K + $^{1}$H, respectively. The apparent worsening of the agreement
between the experimental and calculated $\alpha$-particle spectra (dashed circular
arcs) at larger
angle results from the relatively large low-energy thresholds of the most
backward-angle telescopes. Despite these artifacts, the spectra can be
understood by assuming a sequential evaporative process and successive emission
sources starting with the thermally equilibrated $^{40}$Ca$^{*}$ CN and ending
with the final source characterised by a complete freeze-out of the residual
nucleus. It is clear from this figure that the Galilean-invariant cross section
contours fall on the dashed circular arcs centered at V$_{c.m.}$, as expected
for a complete fusion-evaporation (CF) mechanism followed by isotropic
evaporation. 

\subsection{Exclusive data}

\subsubsection{LCP energy spectra}

The exclusive LCP events are consistent with a CF mechanism (as shown in the
previous analysis of the inclusive LCP data) as they are mainly in coincidence
with ER's with Z = 15-17. Figs.~7 and 8 display the exclusive energy spectra of
$\alpha$ particles detected at the indicated angles (from $\Theta^{LCP}_{lab}$
= +40$^{\circ}$ to $\Theta^{LCP}_{lab}$ = +65$^{\circ}$ (+95$^{\circ}$)), in
coincidence with individual ER's (Z = 15 and 16) identified in the IC detector
located at $\Theta^{ER}_{lab}$ = -15$^{\circ}$ (and -10$^\circ$) in the
$^{28}$Si + $^{12}$C reaction at E$_{lab}$ = 112 MeV and 180 MeV, respectively.
The data taken with the IC's located at more backward angles (larger than
-15$^{\circ}$) are not considered  in the following analysis since the
statistics for fusion-evaporation events are too low. The experimental data are
given by the solid points, with error bars visible when greater than the size
of the points. The spectral shapes of $\alpha$ particles in coincidence with P
(Z=15) ER's, shown in Fig.~8, are very similar to the inclusive energy spectra
of Fig.~5. On the other hand the energy spectra of $\alpha$ particles in
coincidence with S (Z=16) ER's are more complicated as they show other
sub-structures which are superimposed on the ``statistical" Maxwellian shape. 
In Fig.~7 for Z = 15 the high-energy components showing up at the large angles
may arise from $\alpha$, 3p evaporation cascades.

Fig.~9 presents the corresponding $^{28}$Si + $^{12}$C exclusive energy spectra
of protons emitted in coincidence with individual ER's (Z = 15 and 16) at
E$_{lab}$ = 180 MeV. Their spectral shapes are also Maxwellian with the typical
exponential fall-off at high energy, characteristic of a statistical CN decay
process. 

\subsubsection{In-plane angular correlations}

The in-plane angular correlations of $\alpha$ particles and protons (measured
with -115$^{\circ}$ $\leq$ $\Theta^{LCP}_{lab}$ $\leq$ +115$^{\circ}$) in
coincidence with ER's (14 $\leq$ Z $\leq$ 17), produced in the $^{28}$Si +
$^{12}$C reaction, are shown in Figs.~10 and 11 at E$_{lab}$ = 112 MeV and 180
MeV, respectively. The angular correlations are peaked strongly on the opposite
side of the beam direction from the ER detectors which were located at
$\Theta^{ER}_{lab}$ = -15$^{\circ}$ or $\Theta^{ER}_{lab}$ = -10$^{\circ}$ for
the two energies, respectively. The observed peaking of the LCP yields on the
opposite side of the beam from the IC is the result of momentum conservation.
The angular correlations of both the protons and $\alpha$ particles show the
same behavior for all ER's. The solid lines shown in the figures are the
results of statistical-model predictions described in the next Section. 

\subsubsection{Out-of-plane angular correlations}

Fig.~12 displays the out-of-plane angular correlations of $\alpha$ particles
(circles) and protons (triagles) in coincidence with individual ER's detected
at $\Theta_{lab}$ = 10$^\circ$, produced in the $^{28}$Si(180 MeV) + $^{12}$C
reaction. The angular distributions have a behaviour following a typical
exp(-a$sin^2$($\theta_{lab}$))
shape~\cite{Rousseau01a,Beck95,Mahboub96,Mahboub02}, with possibly two
components visible for $\alpha$ particles in coincidence with Z = 14, 15 and
16, plus a broadening of the angular correlations at backward angles. This
broadening effect may result from $\alpha$ particles being able to be emitted
at the beginning or at the end of the decay chain, where the angular momentum
becomes smaller towards the end of the chain. As the protons cannot remove as
much angular momentum as do the $\alpha$ particles the broadening effect is
less significant in the proton angular correlation. The solid lines shown in
the figure are the results of statistical-model predictions for CF and
equilibrium decay using the Monte Carlo evaporation code {\sc
CACARIZO}~\cite{Choudhury84,Majka87,Viesti88}, as discussed in the next
Section. 

\newpage

\section{DISCUSSION}

\subsection{Statistical-model calculations}

The evaporation of light particles from a highly excited CN is a well known
decay process up to very high excitation energies and
spins~\cite{Stokstad85,Cole00,Charity00}. The interpretation of LCP data
requires a careful treatment of the light particle emission properties in the
statistical-model description. Most of the available statistical-model computer
codes~\cite{Sanders99,Matsuse97,Stokstad85,Cole00,Charity00} are based on the
Hauser-Feshbach formalism and are able to follow the CN decay by a cascade of
evaporated LCP's and neutrons. In particular, a detailed analysis of the
exclusive data can be undertaken by the use of Monte Carlo
versions~\cite{Cole00} of some of these statistical-model codes in which the
filtering of the events can reproduce the experimental conditions. The
statistical-model analysis of the present data has been performed using the
Hauser-Feshbach evaporation code {\sc CACARIZO}~\cite{Choudhury84}. {\sc
CACARIZO} is a Monte Carlo version of the statistical-model code {\sc
CASCADE}~\cite{Puhlhofer77}, which has evolved with many modifications and
extensions~\cite{Choudhury84,Majka87,Viesti88} from the original code. In this
program the effective experimental geometry of the {\sc ICARE} detectors is
properly taken into account. It is assumed that a single CN is created with a
well defined excitation energy and angular momentum distribution, and the
de-excitation chain is followed step by step and recorded as an event file. The
generated events are then analyzed using a subsequent filtering code {\sc
ANALYSIS}~\cite{Majka87} in which the locations and the solid angles of all the
{\sc ICARE} telescopes are explicitly specified. This program allows the
determination of the different types of events of interest. Such events can  be
sorted (singles events, coincidence events, etc.) and the corresponding
particle spectra and angular distributions can be created. 

The CN  angular momentum distributions needed as the primary input for the
calculations were specified using the ER critical angular momentum L$_{crit}$
and the diffuseness parameter $\Delta$L. They were taken from ER cross section
data compiled for the $^{28}$Si + $^{12}$C fusion process by Vineyard {\it et 
al.}~\cite{Vineyard93}, without including fission competition. A fixed
value of $\Delta$L = 1$\hbar$ (optimized at low energy by a previous
statistical-model analysis of this reaction~\cite{Alamanos83}) was assumed for
the calculations. It has been checked that the calculated spectra are not
sensitive to slight changes in the critical angular momentum or to explicit
inclusion of the fission competition. The parameter sets used for the
calculations are summarised in Table II. 

The other standard ingredients for statistical-model calculations are the
formulations of the nuclear level densities and of the barrier transmission
probabilities. The transmission coefficients were derived from Optical Model
(OM) calculations using potential parameters of light particle induced
reactions deduced by Wilmore and Hodgson~\cite{Wilmore64}, Perey and
Perey~\cite{Perey63}, and Huizenga~\cite{Huizenga61} for the neutrons, protons
and $\alpha$ particles, respectively. For spin regions where the standard
rotating liquid drop model (RLDM)~\cite{Cohen74} as well as the finite-range
liquid drop model (FRLDM)~\cite{Sierk86} still predict essentially spherical
shapes, these sets of transmission coefficients have been found adequate in the
considered mass region. However, in recent years it has been observed that when
the angular momentum is increased to values for which FRLDM predicts
significant deformations, statistical-model calculations using such standard
parameters cannot always predict satisfactorily the shape of the evaporated
$\alpha$-particle energy spectra 
\cite{Choudhury84,Majka87,Govil87,Fornal88,Viesti88,Fornal89,Larana89,Huizenga89,Fornal91a,Agnihotri93,Govil98,Bandyopadhyay99,Govil00a,Govil00b,Bandy01}.
The calculated average energies of the $\alpha$ particles are found to be much
higher than the corresponding experimental results. Several attempts have been
made to explain this anomaly either by changing the emission barrier or by
using spin-dependent level densities. Adjusting the emission barriers and
corresponding transmission probabilities affects the lower-energy part of the
calculated evaporation spectra. On the other hand the high-energy part of the
spectra depends crucially on the available phase space obtained from the level
densities at high spin. In hot rotating nuclei formed in heavy-ion reactions,
the energy level density at higher angular momentum is spin dependent. The
level density, $\rho(E,J)$, for a given angular momentum $J$ and energy $E$ is
given by the well known Fermi gas expression with equidistant single-particle
levels and a constant level density parameter $a$: 

\begin{equation}
\rho(E,J) = {\frac{(2J+1)}{12}}a^{1/2}
           ({\frac{ \hbar^2}{2 {\cal J}_{eff}}}) ^{3/2}
           {\frac{1}{(E-\Delta-T-E_J)^2} }$\rm exp$(2[a(E-\Delta-T-E_J)]^{1/2})
\label{lev}
\end{equation}
\noindent
where T is the ``nuclear" temperature and $\Delta$ is the pairing
correction~\cite{Dilg73}. The quantity E$_J$ = $\frac{ \hbar^2}{2 {\cal
J}_{eff}}$J(J+1) is the rotational energy, with ${\cal J}_{eff} = {\cal J}_0
\times (1+\delta_1J^2+\delta_2J^4)$ being the effective moment of inertia,
where ${\cal J}_0$ at high excitation energy and high angular momentum is
considered to be the rigid body moment of inertia and $\delta_1$ and $\delta_2$
are the ``deformability parameters" 
\cite{Choudhury84,Majka87,Govil87,Viesti88,Huizenga89,Fornal91a,Agnihotri93,Govil98,Bandyopadhyay99,Govil00a,Govil00b,Bandy01}.

The level density parameter is constant and is set equal to $a$ = A/8
MeV$^{-1}$, a value which is in agreement with previous works 
\cite{Fornal91b,Shlomo91,Toke81}. In principle, the value of $a$ may be
affected by dynamical deformation: rotation induces rearrangement of the
single-particle level scheme and the altered nuclear surface area
~\cite{Huizenga89} affects the macroscopic energy of the system. The $a$
parameter becomes more important when the nuclear deformation
increases~\cite{Toke81}. However, in the present work we assume a constant
value and rather introduce deformation effects through the deformability
parameters. A constant value of $a$ = A/8 is in agreement with various
authors~\cite{Huizenga89,Fornal91b}, as well as with theoretical studies by
Shlomo and Natowitz~\cite{Shlomo91}, by T\"oke and Swiatecki~\cite{Toke81},
and with experimental results obtained very recently in the A$_{CN}$ = 60 mass
region~\cite{Janker99}. 

Rather than to adjust the spin-dependence of the moment of inertia as done
here, with effective emission barriers being unchanged, an alternative method
to mock up nuclear deformation is to vary the radius paramater r$_{0}$ (from 
RLDM~\cite{Cohen74} parameters proposed by Myers and Swiatecki~\cite{Myers66})
of the rigid-body moment of inertia ${\cal J}_0$, as discussed recently for
very light nuclear systems such as $^{31}$P~\cite{Bandyopadhyay99,Bandy01}.
This will affect the transmission coefficients by lowering the effective
emission barrier~\cite{Huizenga89}. However, a crude increase of the radius
parameter (up to 25$\%$) in the OM transmission coefficients has been
shown~\cite{Huizenga89} to enhance the intensity of the high-energy $\alpha$
particles in the spectra. Although for hot heavy nuclei at high excitation
energy a lowering of the $\alpha$-particle emission barriers has to be taken
into account~\cite{Viesti01}, this artifice has been extensively criticized in
the past for lighter systems~\cite{Huizenga89}. In fact this would result in a
poorer reproduction of the present $^{28}$Si + $^{12}$C data. Therefore no
attempt was made to modify the transmission coefficients since it has been
shown that the effective barrier heights are fairly insensitive to the nuclear
deformation~\cite{Huizenga89}. On the other hand, by changing the deformability
parameters $\delta_1$ and $\delta_2$ one can simulate the spin-dependent level 
density~\cite{Govil87,Viesti88,Huizenga89,Agnihotri93} associated with a larger
nuclear deformation, and thus reproduce the experimental data in a much better
way. 

\subsection{Deformation effects in $^{40}$Ca}

In the present analysis, following the procedure proposed by Huizenga {\it et
al.}~\cite{Huizenga89}, we empirically modify the phase space open to
statistical decay by lowering the Yrast line with adjustment of the
deformability parameters so as to fit the available experimental
data~\cite{Govil87,Viesti88}. We take into account the fact that the
deformation should be attenuated during the subsequent emission processes:
i.e., there is a readjustment of shape of the nascent final nucleus and a
change of collective to intrinsic excitation during the particle-evaporation
process. A similar analysis was suggested earlier by Blann and
Komoto~\cite{Blann81}, but with the assumption that the deformation is a frozen
degree of freedom through the decay chain. Dynamical effects related to the
shape relaxation during the de-excitation process have been incorporated into
statistical-model codes~\cite{Fornal89,Fornal91a}. For the CACARIZO
calculations done here, it is assumed that memory of formation details are lost
after each step, with only the conserved quantities such as total energy and
spin preserved during the decay sequence. The {\sc CACARIZO} calculations have
been performed using two sets of input parameters: the first one with standard
liquid drop parameters (parameter {\bf set A}), consistent with the deformation
of RLDM~\cite{Cohen74} and of FRLDM with finite-range corrections of
Sierk~\cite{Sierk86}, and the second one with larger values for the
deformability parameters
\cite{Bhattacharya99,Bhattacharya02,Rousseau00,Beck00,Bhattacharya01}
(parameter {\bf set B}) which are listed in Table \ref{table3}. 
 
The dashed lines in Fig.~5 show the predictions of {\sc CACARIZO} for $^{28}$Si
+ $^{12}$C at E$_{lab}$ = 180 MeV using the parameter {\bf set A} consistent
with FRLDM deformation~\cite{Sierk86}. It is clear that the average energies of
the measured $\alpha$-particle inclusive spectra are lower than those predicted
by these statistical-model calculations. The solid lines of Fig.~5 show the
predictions of {\sc CACARIZO} using the increased values of the spin deformation
parameters (see parameter {\bf set B} given in Table II), and the agreement is
considerably improved. 

The exclusive energy spectra of the $\alpha$ particles in coincidence with
individual ER's (Z = 15 and Z = 16) are displayed in Figs.~7 and 8 for the two
bombarding energies E$_{lab}$ =  112 MeV and 180 MeV, respectively. It can be
observed that the spectra in coincidence with the P residues are well
reproduced by using the deformation
effects~\cite{Bhattacharya99,Rousseau00,Beck00,Bhattacharya01}. The solid lines
in Figs.~7 and 8 show the predictions of {\sc CACARIZO} using the parameter
{\bf set B} with $\delta_1$ = 2.5 x 10$^{-4}$ and $\delta_2$ = 5.0 x 10$^{-7}$
chosen to reproduce the data consistently at the two bombarding energies. On
the other hand, by using the standard liquid drop deformability parameter {\bf
set A} with no extra deformation (i.e. with small values of $\delta _1$ and
$\delta_2$), the calculated average energies from the exclusive
$\alpha$-particles spectra (not shown in Figs.~7 and 8) are, as found for the
inclusive data, lower than those predicted \cite{Rousseau01a} by the
statistical model. In this case the {\sc CACARIZO} parameters are similar to
the standard parameters used in a previous study of the 130 MeV $^{16}$O +
$^{24}$Mg reaction \cite{Fornal91b}, with the use of the angular momentum
dependent level densities. 

The exclusive energy spectra of $\alpha$ particles measured in coincidence with
individual S ER's, which are shown in Figs.~7 and 8 for E$_{lab}$ = 112 MeV and
180 MeV, respectively, are quite interesting. A large difference can be
noticed when comparing the energy spectra associated with S residues and those
associated with P ER's~\cite{Beck00}. The latter are reasonably well reproduced
by the {\sc CACARIZO} calculations, whereas the model does not predict the
shape of the spectra obtained in coincidence with S residues at backward angles
($\theta_{\alpha}\geq 40^{\circ}$ at E$_{lab}$ = 112 MeV and
$\theta_{\alpha}\geq 70^{\circ}$ at E$_{lab}$ = 180 MeV). Additional
non-evaporative components appear to be significant in this case. This is
consistent with the discrepancies also observed at backward angles in the
in-plane angular correlations of Fig.~10. In the following section we will
discuss the hypothesis that the discrepancy occurs because a strong cluster
transfer reaction mechanism leads to significant $^{8}$Be production, with the
$^{8}$Be subsequently breaking up into two $\alpha$ particles. 

As shown in Fig.~9 {\sc CACARIZO} calculations are also capable to reproduce
the shape of exclusive proton spectra for the $^{28}$Si(180 MeV) + $^{12}$C
reaction. Compared to the $\alpha$ particles, it may be mentioned that the
energy spectra of the protons do not shift as significantly as the
spin-dependent parametrization of the moment of inertia is introduced. The
statistical-model results using the two different parameter sets reproduce
equally well the experimental velocity spectra and angular correlations. The
statistical-model calculations displayed for protons in Fig.~9 have been
performed with parameter {\bf set B} (solid lines) including the deformation
effects (calculations with parameter {\bf set A} are not displayed). 

In order to better determine the magnitude of the influence of deformation
effects in the CN and the residual nuclei which are suggested by our choice of
statistical-model approach, we have proposed a very simple
procedure~\cite{Bhattacharya02,Beck95,Mahboub96,Mahboub02}. The effective
moment of inertia is 
expressed as 
${\cal J}_{eff}$ = ${\frac{2}{5}}$MR$^{2}$ = ${\frac{1}{5}}$M(b$^{2}$+a$^{2}$)
with the volume conservation condition: V = ${\frac{4}{3}}\pi$abc, 
where b and a are the major and minor axis, and c is the rotational axis
of the CN.
In the case of an oblate shape a = b and ${\cal J}_{eff}$ =
${\frac{2}{5}}$Ma$^{2}$ 
and V = ${\frac{4}{3}}\pi$a$^{2}$c. The axis ratio is equal to 
$\delta$ = a/c = (1+$\delta_1$J$^{2}$+$\delta_2$J$^{4}$)$^{3/2}$.
In the case of a prolate shape a = c and ${\cal J}_0$ =
${\frac{1}{5}}$M(b$^{2}$+a$^{2}$) and  V = ${\frac{4}{3}}\pi$a$^{2}$b. We
obtain the equation: 1+(3-$\gamma$)x+3x$^{2}$+x$^{3}$ = 0 with x = $\left(
{\frac{b}{a}}\right)^{2}$ = ${\delta}^{2}$ and $\gamma$ =
8(1+$\delta_1$J$^{2}$+$\delta_2$J$^{4}$)$^{3}$.
The quadrupole deformation parameter $\beta$ is equal to $\beta$ =
${\frac{1}{\sqrt{5\pi}}}({\frac{4}{3}}\delta+{\frac{2}{3}}\delta^{2}+{\frac{2}{3}}\delta^{3}+{\frac{11}{18}}\delta^{4})$.\\

The effects of the Yrast line lowering (increase of the level density)
due to the nuclear deformation and the variation of the deformation parameter
$\beta$ can be quantitatively discussed using the values in Table \ref{table3}
for several reactions. The values of the minor
to major axis ratio b/a and of the deformation parameter $\beta$ have been
extracted from the fitted deformability parameters by assuming either a
symmetric prolate shape or a symmetric oblate shape, respectively, with sharp
surfaces~\cite{Huizenga89}. In our analysis we have chosen to follow the
procedure proposed by Huizenga {\it et al.}~\cite{Huizenga89} and successfully
used for the $^{28}$Si + $^{28}$Si data~\cite{Bhattacharya02}. No attempt
to modify the parametrization of OM transmission coefficients has been
undertaken since it has been shown that the effective barrier heights are
fairly insensitive to the nuclear deformation~\cite{Huizenga89}. It is
interesting to note that the deformation found necessary to reproduce the
$^{28}$Si + $^{12}$C reaction results is smaller than the deformation
introduced by the deformability parameter used by Kildir {\it et
al.}~\cite{Kildir95}, who also change the transmission coefficients. 

The in-plane angular correlations of the $\alpha$ particles (circles) and the
protons (triangles) in coincidence with individual ER's at both energies
E$_{lab}$ = 112 MeV and 180 MeV are shown in Figs.~10 and 11, respectively. The
solid lines shown in the figures are the results of statistical-model
predictions for CF and equilibrium decay using the evaporation code {\sc
CACARIZO}. It can be observed in Fig.~10 that for E$_{lab}$ = 112 MeV the
experimental angular correlations are well reproduced by the evaporation
calculations for the data at the opposite side from the ER detector, and this
is true for correlations with both S and P ER's. However the calculations fail
to predict the experimental data at the same side as the ER detector. The
question of these large yields measured at negative angles remains
open~\cite{Bhattacharya02}. Similarly the {\sc CACARIZO} calculations reproduce
in Fig.~11 the in-plane angular correlations of $\alpha$ particles (circles)
and protons in coincidence with all ER's, at E$_{lab}$ = 180 MeV, for the data
on the opposite side from the ER detector. They are also able to describe the
in-plane angular correlations of protons in coincidence with individual Z = 14
and Z = 15 on both sides of the beam. However, the excess of yields observed at
backward angles ($\Theta_{lab}$ = +50$^\circ$ to +90$^\circ$) for $\alpha$
particles in coincidence with S may indicate the occurence of a non-evaporative
process, possibly of a binary nature. 

The solid lines shown in Fig.~12 for the out-of-plane angular correlations are
the results of {\sc CACARIZO} statistical-model predictions. Once again it can
be observed that the statistical-model calculations are able to reproduce the
proton coincidences well, but they fail to describe the $\alpha$-Cl
coincidences and the large yields found in coincidence with P and Si ER's in
the most forward direction. The reason why the experimental anisotropy factor
is not well reproduced by the calculations is not well understood. The angular
momentum dependence has been tested by performing calculations with two
different angular-momentum windows : 10-20$\hbar$ and 20-30$\hbar$. Whereas for
protons the anisotropy is almost constant with the L-window, for the $\alpha$
particles the anisotropy is strongly depending of the chosen L-window.
Nevertheless the flat behavior shown around 0$^\circ$ is present for the two
particle species. The problem of this discrepancy in the out-of-plane angular
correlations may need to use a more complete formulation for the angular
distribution of the LP which is treated in a semi-classical way. 

The main conclusion that can be drawn from the above analysis indicate that the
shape of the $^{40}$Ca nuclear system is expected to be elongated by rather
significant deformation effects during the evaporative processes of $\alpha$
particles. This strong deformation is compatible with the previous analysis of
$^{28}$Si($^{12}$C,2$\alpha$)$^{32}$S$_{g.s.}$ angular correlation
data~\cite{Alamanos83}. The extent to which these effects can be reasonably
well quantified is dependent on the degree of complexity of the experimental
device and, in particular, on the power of the coincidence trigger. It is of
particular interest to note that the value of $\beta$ $\approx$ 0.5 found for
the quadrupole deformation parameter of $^{28}$Si + $^{12}$C (see Table III)
might be connected with the recent observation of SD bands in the doubly-magic
$^{40}$Ca nucleus by standard $\gamma$-ray spectroscopy
methods~\cite{Ideguchi01}. Correlating large deformations in the hot CN with
the presence of SD bands in $^{40}$Ca is obviously not straightforward, since
the deformation deduced from the LCP data is the average, while SD bands are
one of the possible configurations. We made the same discussion with the
possible comparison between LCP results for the $^{28}$Si + $^{28}$Si
reaction~\cite{Bhattacharya02} and $\gamma$-ray data displaying very deformed
bands in the doubly magic $^{56}$Ni nucleus~\cite{Rudolph99}. 

\subsection{Non-statistical $^{8}$Be cluster emission}

It has been shown from the analysis of Figs.~7 and 8 with {\sc CACARIZO} that
additional non-statistical components appear to be significant at both
bomabarding energies E$_{lab}$ = 112 MeV and 180 MeV. However no evidence was
found for additional processes at the lower bombarding energies E$_{lab}$ = 70
MeV~\cite{Alamanos83} and 87 MeV~\cite{Ost80}. To better understand the origin
of these components, $\alpha$ particle energies are plotted in Fig.~13 against
the energies of the S residues detected at $\Theta_{S}$ = -10$^\circ$ for the
$^{28}$Si(180 MeV) + $^{12}$C reaction for a number of $\alpha$-particle
emission angles. With increasing $\alpha$-particle angles an increase of the
energy of S residues and a decrease of the $\alpha$ energy is observed which is
consistent with kinematics. At $\Theta_{\alpha}$ = +40$^\circ$, +45$^\circ$,
and +50$^\circ$ the bulk of events in Fig.~13 are of a statistical origin, and
consistent with {\sc CACARIZO} calculations, as demonstrated in Fig.~14 (for
$\Theta_{\alpha}$ = +40$^\circ$). Another statistical-model code PACE
2~\cite{Gavron80} gives similar predictions. The calculations suggest that
these $\alpha$ particles result from a cascade of a single $\alpha$, two
protons, and x neutrons rather than a 2-$\alpha$,xn evaporation process. For
larger angles, the two branches, corresponding to the contours labelled 1 and
2, although lying outside the ``statistical evaporation region'', still
correspond to an evaporation process as shown by the {\sc CACARIZO}
calculations displayed in Fig.~14 for $\Theta_{\alpha}$ = +40$^\circ$ and
+70$^\circ$. These two branches 1 and 2 correspond to a 2-$\alpha$
fusion-evaporation channel with both the $\alpha$ particles emitted
respectively at backward and forward angles in the center of mass. However, at
more backward angles other contributions, corresponding to the contours
labelled 3 and 4, appear more and more significantly as shown for instance in
Fig.~14 for +70$^\circ$. The corresponding ``folding angles'' are compatible
with the two-body kinematics required for the $^{32}$S + $^{8}$Be binary
exit-channel. In contrast, the energy correlations for the $\alpha$ particles
in coincidence with Cl and P residues (not shown) do not exhibit any of these
two-body branches 3 and 4 and, thus, the ``statistical evaporation region'' is
consistent with the {\sc CACARIZO} predictions, for all the measured angles. 

Although in principle the identification of the $^{8}$Be cluster requires the
coincident detection and mass identification of both decaying $\alpha$
particles~\cite{Mcdonald92}, a kinematic reconstruction of this decay process
is still possible from the present coincident data to determine the excitation
energy of the $^{8}$Be nucleus by assuming a two-body $^{32}$S + $^{8}$Be$^{*}$
process. On the left side of Fig.~15 the excitation energy of $^{8}$Be is
presented for the contributions labelled 2, 3, and 4 (Fig.~14) at the indicated
$\Theta_{\alpha}$ angles. From $\Theta_{\alpha}$ = 70$^\circ$ to
$\Theta_{\alpha}$ = 85$^\circ$ the strongest peak appears with a very narrow
width. This large component, which corresponds to the contribution of the
contour 4 visible in Fig.~14, is centered at the energy of the ground state of
$^{8}$Be (the relative energy of the two $\alpha$ particles of the $^{8}$Be
breaking up in flight is 92 keV) and displayed as the squared part of the left
pannel of Fig.~15. From $\Theta_{\alpha}$ = 55$^\circ$ to $\Theta_{\alpha}$ =
95 $^\circ$ the main bulk of the yields from contour 3 is centered at around
E$^{*}$ = 3.1 MeV with an experimental width of approximately 1.5 MeV, which
values correspond well to the known energy (E$^{*}$ = 3.04 MeV) and width
($\Gamma$ = 1500 keV) of the first 2$^+$ excited level of
$^{8}$Be~\cite{Ajzenberg88}. The short-lived $^{8}$Be 4$^+$ excited level at
E$^{*}$ = 11.4 MeV~\cite{Arena95} is not clearly observed due to its very broad
width ($\Gamma$ = 3.7 MeV) and the significant $\alpha$-statistical background
arising from the contribution of the contour 2. For the same reasons it is
hazardous to assign the bumps around 15 MeV to the known 2$^+$
doublet~\cite{Ajzenberg88} at E$^{*}$ = 16.6 and 16.9 MeV. 

The right pannels of the Fig.~15 display the reconstructed excitation energy
spectra of the S binary fragments measured at $\Theta_{S}$ = -10$^\circ$ in
coincidence with $\alpha$ particles detected at the indicated $\Theta_{\alpha}$
angles by gating either on the groud state (g.s.) contour 4 (upper panel) or
the 2$^+$ state contour 3 (lower panel). We have performed fusion-fission
calculations (not shown), using the Extended Hauser-Feshbach Method
\cite{Matsuse97}. They fail to reproduce both the excitation energies of the
$^{32}$S fragments, and the yields from the contributions 3 and 4
\cite{Rousseau00}. These contributions might result from a faster binary
process governed by the $\alpha$-transfer reaction mechanism $^{28}$Si +
$^{12}$C $\rightarrow$ $^{32}$S$^{*}$ + $^{8}$Be, as proposed by Morgenstern
{\it et al.}~\cite{Morgenstern86}. This conclusion is in agreement with
previous inclusive results published in Ref.~\cite{Arena94}. In the
cluster-transfer picture~\cite{Morgenstern86} the reaction is characterised by
a ``Q-value'' window centered at the so-called ``Q-optimum'', which value can
be estimated semi-classically by Q$_{opt}$ =
(Z$_{3}$Z$_{4}$/Z$_{1}$Z$_{2}$-1)E$_i^{c.m.}$, where the indices 1,2 and 3,4
indicate the entrance (i) and exit channel, respectively. The corresponding
excitation energy E$^{*}$ = Q$_{gg}$ - Q$_{opt}$, where Q$_{gg}$ is the
ground-state Q-value of the reaction. In this case the expected excitation
energy in the $^{32}$S nuclei is equal to 12.9 MeV. The right hand side of
Fig.~15 represents the calculated excitation energy of $^{32}$S in coincidence
with the g.s., and with the first 2$^{+}$ (E$_x$ = 3.04 MeV) excited state of
${^8}$Be, respectively. The dashed lines correspond to E$^{*}$ = 12.9 MeV, the
energy expected for $\alpha$-transfer reaction mechanisms. In both cases the
excitation energies of $^{32}$S are consistent with these values. In the same
way we can also have a $^{8}$Be-transfer reaction mechanism \cite{Arena94}
$^{28}$Si + $^{12}$C $\rightarrow$ $^{36}$Ar$^{*}$ + $\alpha$. In this case the
$^{36}$Ar$^{*}$ ejectile has enough excitation energy to emit either one proton
or one $\alpha$ particle. This type of ``transferlike'' reaction can explain
the disagreement observed in Fig.~10 between data and {\sc CACARIZO}
calculation for the in-plane angular correlation between $\alpha$ particles and
Cl residues. 

\newpage

\section{CONCLUSION}

The possible occurence of highly deformed configurations in the $^{40}$Ca
dinuclear system has been investigated by using the {\sc ICARE}
charged-particle multidetector array at the {\sc VIVITRON} Tandem facility of
the IReS Strasbourg. The properties of the emitted LCP's in the $^{28}$Si +
$^{12}$C reaction, have been analysed at two bombarding energies E$_{lab}$ =
112 MeV and 180 MeV, and compared with a statistical model that was adopted to
calculate evaporation spectra and angular distributions for deformed nuclei. A
Monte Carlo technique has been employed in the framework of the well documented
Hauser-Feshbach code {\sc CACARIZO}. The measured observables such as velocity
distributions, energy spectra, in-plane and out-of-plane angular correlations
are all reasonably well described by the Monte Carlo calculations when they
include spin-dependent level densities, and indicate that effects due to
differences in nuclear shapes are large enough to be observed. The magnitude of
the adjustements in the Yrast line suggests significant deformation effects at
high spin for the $^{40}$Ca dinuclear system comparable to recent $\gamma$-ray
spectroscopy data for the $^{40}$Ca nucleus at much lower
spins~\cite{Ideguchi01}. The extent to which the resonant
behaviour~\cite{Pocanic85,Beck94} is responsible to the observed nuclear
deformation is still an open question. A non-statistical binary component is
found in the $\alpha$-particle energy spectra measured in coincidence with S
residues that is attributed to the cluster decay of unbound $^8$Be nuclei,
produced through the $\alpha$-cluster-transfer reaction $^{28}$Si + $^{12}$C
$\rightarrow$ $^{32}$S + $^{8}$Be~\cite{Arena94}. This new type of
``transferlike'' mechanism, which does not appear at lower bombarding
energies~\cite{Alamanos83,Ost80} below threshold, allows to populate some N = Z
nuclei with a well defined excitation energy. Therefore, sophisticated
particle-$\gamma$ experiments (see Refs.
\cite{Nouicer99,Beck01,Thummerer00,Thummerer01} for instance) using {sc
EUROBALL IV} and/or {sc GAMMASPHERE} should be performed in the very near
future in order to well define and understand what are the best types of
reaction which can populate significantly the superdeformed bands discovered
and/or predicted in this mass region~\cite{Svensson00,Svensson01,Ideguchi01}.

\newpage

\centerline{\bf ACKNOWLEDGMENTS }

\vskip 0.8cm

\noindent
This paper is based upon the Ph.D.~thesis of M. Rousseau, Universit\'e Louis
Pasteur, Strasbourg, 2000. The authors would like to thank the staff of the
{\sc VIVITRON} for providing us with good stable beams, M.A. Saettel for
preparing the targets, and J. Devin and C. Fuchs for their excellent support in
carrying out these experiments. Particular appreciation to M.A. Saettel for
preparing the targets, and to J.P.~Stockert and A.~Pape for assistance during
their RBS measurements. We also whish to thank N. Rowley and W. Catford for
useful discussions and for a careful reading of the manuscript. One of us
(M.R.) would like to acknowledge the Conseil R\'egional d'Alsace for the
financial support of his Ph.D.~thesis work. Parts of this work has also been
done in collaboration with C.E. during his summer stay at the IReS with a JANUS
Grant of IN2P3. This work was sponsored by the French CNRS/IN2P3, and partially
by the CNRS/NSF and CNRS/CNPq collaboration programs.

\hspace{-2cm}\renewcommand{\baselinestretch}{2.}
\begin{table}[hptb]
\begin{tabular}{|c|c|c|}
\hline
 &\multicolumn{2}{c|}{$\sigma \pm \Delta\sigma$ (mb)} \\
\hline
Z & $^{28}$Si(112 MeV) + $^{12}$C  & $^{28}$Si(180 MeV) + $^{12}$C    \\ 
\hline
5 & -                 &  8.4  $\pm$ 1.1 \\
6 & 20.1 $\pm$ 1.4  &  58.8 $\pm$ 6.6 \\
7 &  0.5 $\pm$ 0.04  &  11.8 $\pm$ 1.4 \\
8 &  3.3 $\pm$ 0.3  &  23.1 $\pm$ 3.1 \\
9 & -                 &  4.7  $\pm$ 0.7 \\
10 &  0.8 $\pm$ 0.12 &  18.9 $\pm$ 4.0 \\
11 & -                &  15.9 $\pm$ 3.4 \\
\hline
\end{tabular}
\renewcommand{\baselinestretch}{1.}
\caption{\label{table1}\textit{\sl Experimental binary fragments cross sections 
obtained in the $^{28}$Si + $^{12}$C reaction at E$_{lab}$ = 112 MeV and 180
MeV.}} 
\end{table}
\renewcommand{\baselinestretch}{1.}


{\footnotesize
\begin{table}[htpb]
\begin{tabular}{|l|}
\hline
\hline
\centerline{\bf Angular-momentum distribution in CN}\\
\\
Critical angular momenta $L_{cr}$ = 21 (E$_{lab}$ = 112 MeV) and 27$\hbar$
(E$_{lab}$ = 180 MeV).\\
Diffuseness parameter $\Delta L = 1.0\hbar$.\\
\hline
\centerline{\bf OM potentials of the emitted LCP and neutrons}\\
\\
(1) Neutrons: Wilmore and Hodgson~\cite{Wilmore64}\\
(2) Protons: Perey and Perey~\cite{Perey63}\\
(3) $\alpha$ particles: Huizenga and Igo~\cite{Huizenga61}\\
(4) Multiply factor of the OM radius: RFACT = 1\\
\hline
\centerline{\parbox{14cm}{\bf Level-density parameters at low excitation:
(E$^* \leq$ 10 MeV)}}\\
\\
(1) \parbox{15cm}{Fermi-gas level-density formula with empirical level-density
parameters from Dilg {\it et al.}~\cite{Dilg73}}\\
(2) Effective moment of inertia $\Im$ = IFACT $\Im_{rigid}$ with IFACT = 1.\\
\hline
\centerline{\parbox{14cm}{\bf Level-density parameters at high excitation:
(E$^* \geq$ 20 MeV)}}\\
\\
(1) \parbox{15cm}{ Fermi-gas level-density formula with parameters from RLDM
(Myers and Swiatecki~\cite{Myers66})}\\
(2) Level-density parameter: $a$ = A/8 MeV$^{-1}$\\
\hline
\centerline{\bf Yrast line}\\
\\
- Parameter set {\bf A}: FRLDM (Sierk~\cite{Sierk86})
\\
- Parameter set {\bf B}: $\Im = \Im_{sphere} (1 + \delta_1 J^2 +
\delta_2 J^4)$ with $\delta_1$ = 2.5 10$^{-4}$ et $\delta_2$ = 5.0 10$^{-7}$\\
\hline
\centerline{{\bf $\gamma$-ray width} (in Weisskopf units)}\\
\\
(1) E1: B(E1) = 0.001\\
(2) M1: B(M1) = 0.01\\
(3) E2: B(E2) = 5.0\\
\hline

\end{tabular}

\caption{\label{table2}\textit{\sl Parameter sets used in the CACARIZO
calculations for the $^{28}$Si + $^{12}$C reaction at E$_{lab}$ = 112 and 180
MeV.}}

\end{table}}

\renewcommand{\baselinestretch}{1.2}
\begin{table}[h!]
\hspace{-2cm}
\begin{tabular}{|c|c|c|c|c|c|c|c|c|}
\hline
Reaction & C.N. & Energy (MeV)& $L_{cr}$ ($\hbar$) & $\delta_1$ & $\delta_2$ &
b/a & $\beta$ & Reference\\ 
\hline
$^{28}$Si + $^{12}$C & $^{40}$Ca  & 112 & 21 & 2.5$\cdot$10$^{-4}$ &
5.0$\cdot$10$^{-7}$ & 1.3/1.4 & -0.46/0.47 & This work\\ 
\hline
$^{28}$Si + $^{12}$C & $^{40}$Ca  &  150 & 26 & 6.5$\cdot$10$^{-4}$ &
3.3$\cdot$10$^{-7}$ & 2.0/2.0 & -0.53/0.53 & \cite{Kildir95} \\
\hline
$^{28}$Si + $^{12}$C & $^{40}$Ca  & 180 & 27 & 2.5$\cdot$10$^{-4}$ &
5.0$\cdot$10$^{-7}$ & 1.7/1.8 & -0.51/0.51 & This work\\ 
\hline
$^{28}$Si + $^{27}$Al & $^{55}$Co & 150 & 42 & 1.8$\cdot$10$^{-4}$ &
1.8$\cdot$10$^{-7}$ & 1.2/1.3 & -0.44/0.46 & \cite{Agnihotri93}\\
\hline
$^{28}$Si + $^{28}$Si & $^{56}$Ni & 112 & 34 & 1.2$\cdot$10$^{-4}$ &
1.1$\cdot$10$^{-7}$ & 1.5/1.6 & -0.48/0.49 & \cite{Bhattacharya02}\\ 
\hline
$^{28}$Si + $^{28}$Si & $^{56}$Ni & 180 & 37 & 1.2$\cdot$10$^{-4}$ &
1.1$\cdot$10$^{-7}$ & 1.6/1.7 & -0.49/0.50 & \cite{Rousseau01a}\\
\hline
$^{30}$Si + $^{30}$Si & $^{60}$Ni & 120 & 34 & 1.2$\cdot$10$^{-4}$ &
1.1$\cdot$10$^{-7}$ & 1.6/1.7 & -0.49/0.50 & \cite{Bhattacharya02}\\ 
\hline
$^{35}$Cl + $^{24}$Mg & $^{59}$Cu & 260 & 37 & 1.1$\cdot$10$^{-4}$ &
1.3$\cdot$10$^{-7}$ & 1.6/1.7 & -0.50/0.51 & \cite{Mahboub02}\\  
\hline
$^{32}$S + $^{27}$Al & $^{59}$Cu & 100-150 & 27-42 & 1.3$\cdot$10$^{-4}$
& 1.2$\cdot$10$^{-7}$ & 2.0/2.0 & -0.46/0.53 & \cite{Huizenga89}\\  
\hline
$^{16}$O + $^{54}$Fe & $^{70}$Se & 110 & 34 & 2.5$\cdot$10$^{-5}$ &
3.0$\cdot$10$^{-8}$ & 1.2/1.3 & -0.45/0.46 & \cite{Govil00a}\\
\end{tabular}
\renewcommand{\baselinestretch}{1.}
\caption{\label{table3}\textit{ \sl Typical quantities of the evaporation
calculations performed using the statistical-model code {\sc CACARIZO}. The
deformability parameters are taken either from the parameter {\bf set B} (see
Table II) for the systems studied in the present work or from similar fitting
procedures for the other systems studied in the literature. The minor to major
axis ratios b/a and the quadrupole deformation $\beta_2$ values (for a
symmetric oblate shape and a symmetric prolate shape, respectively) have been
deduced from equations discussed in the text. Note that the $\beta$ values and
b/a ratio given for $^{32}$S + $^{27}$Al have been deduced assuming the $L_{cr}$ as
extracted at the highest bombarding energy. }} 
\end{table}

\begin{figure}
Figure 1: Experimental C (solid squares), N (solid triangles), and O (solid
circles) cross sections measured in the $^{28}$Si + $^{12}$C reaction
\cite{Shapira82,Shapira84} as compared to the calculations (dotted curves)
performed with the equilibrium model of orbiting \cite{Shivakumar87}. The
solid curves are the predictions of the transition-state model
\cite{Sanders99}. The open squares, triangles and circles are the present data
of the C, N, and O fully-damped yields with error bars smaller than the size
of the symbols. The full diamonds correspond to ER cross sections quoted in
Refs.~\cite{Gary82,Lesko82,Nagashima82,Harmon86,Harmon88,Vineyard93,Arena94}. 
\end{figure}

\begin{figure}
Figure 2:  Energy spectra for the C and O fragments measured at $\Theta_{lab}$
= 15$^\circ$ for the $^{28}$Si + $^{12}$C reaction at E$_{lab}$ = 112 MeV. The
peak assignments are discussed in the text. 
\end{figure}

\begin{figure}
Figure 3: Angular distributions (open symbols) of heavy fragments (C, N, O, and
Ne) measured in the $^{28}$Si + $^{12}$C reaction at E$_{lab}$ = 112 MeV. The
full symbols are the C, N and O data taken from Ref.~\cite{Shapira82,Shapira84}
at E$_{lab}$ = 115 MeV. The Ne angular distribution has been scaled down by a
factor 10 for the sake of clarity. The curves correspond to
1/sin$\theta_{c.m.}$ functions. 
\end{figure}
        
\begin{figure}
Figure 4: Angular distributions of heavy fragments (Z = 5 to 11) measured in
the $^{28}$Si + $^{12}$C reaction at E$_{lab}$ = 180 MeV. The N and Na angular
distributions have been scaled down by a factor 10 for the sake of clarity. The
curves correspond to 1/sin$\theta_{c.m.}$ functions. 
\end{figure}

\begin{figure}
Figure 5:  Inclusive energy spectra of $\alpha$ particles measured in the
$^{28}$Si + $^{12}$C reaction at E$_{lab}$ = 180 MeV between
$\Theta^{LCP}_{lab}$ = 30$^\circ$ and 55$^\circ$. The experimental data are
shown by the solid points with error bars visible when greater than the size of
the points. The solid and dashed lines are statistical-model calculations
discussed in the text. 
\end{figure}

\begin{figure}
Figure 6: Two-dimensional scatter plots of Galilean-invariant cross sections
(d$^{2}\sigma$/d$\Omega$dE)p$^{-1}$c$^{-1}$ of inclusive $\alpha$ particles
(left side) and protons (right side) measured in the
(V$_{\perp}$,V$_{\parallel}$) plane for the $^{28}$Si + $^{12}$C reaction at
E$_{lab}$ = 112 MeV (up) and 180 MeV (down). The experimental detector
thresholds are drawn along the laboratory angles of each telescope. The
circular arcs are centered on the velocity of the center of mass. 
\end{figure} 

\begin{figure}
Figure 7: Exclusive energy spectra of $\alpha$ particles emitted at the angles
+40$^{\circ} < \Theta^{LCP}_{lab} < ~$+65$^{\circ}$, in coincidence with
individual P and S ER's detected at -15$^\circ$ in the $^{28}$Si + $^{12}$C
reaction at E$_{lab}$ = 112 MeV. The experimental data are given by the solid
points with error bars visible when greater than the size of the points. The
solid lines are statistical-model calculations discussed in the text. 
\end{figure}
        
\begin{figure}
Figure 8: Exclusive energy spectra of $\alpha$ particles emitted at the angles
+40$^{\circ} < \Theta^{LCP}_{lab} < ~$+95$^{\circ}$, in coincidence with
individual P and S ER's detected at -10$^\circ$ in the $^{28}$Si + $^{12}$C
reaction at E$_{lab}$ = 180 MeV. The experimental data are given by the solid
points with error bars visible when greater than the size of the points. The
solid lines are statistical-model calculations discussed in the text. 
\end{figure}
        
\begin{figure}
Figure 9: Exclusive energy spectra of protons emitted at the angles
+40$^{\circ} < \Theta^{LCP}_{lab} < $+70$^{\circ}$, in coincidence with
individual P and S ER's detected at -10$^\circ$, at the indicated laboratory
angles, in the $^{28}$Si(180 MeV) + $^{12}$C reaction. The solid lines are
statistical-model calculations discussed in the text. 
\end{figure}

\begin{figure}
Figure 10: In-plane angular correlations of coincident $\alpha$ particles
(circles) and protons (triangles) measured in the $^{28}$Si + $^{12}$C reaction
at E$_{lab}$ = 112 MeV. The proton correlations have been multiplied by a
factor 10$^{-3}$ for the sake of clarity. The arrow indicates the position of
the IC detector at $\Theta_{lab}$= -15$^\circ$. On the abscissa, the positive
angle refer to the opposite side of the beam from the direction of the ER
detected in IC. The solid lines correspond to statistical-model calculations
discussed in the text. 
\end{figure}

\begin{figure}
Figure 11: In-plane angular correlations of coincident $\alpha$ particles
(circles) and protons (triangles) measured in the $^{28}$Si + $^{12}$C reaction
at E$_{lab}$ = 180 MeV. The proton correlations have been multiplied by a
factor 10$^{-2}$ for the sake of clarity. The arrow indicates the position of
the IC detector at $\Theta_{lab}$= -10$^\circ$. On the abscissa, the positive
angle refer to the opposite side of the beam from the direction of the ER
detected in IC. The solid lines correspond to statistical-model calculations
discussed in the text. 
\end{figure} 

\begin{figure}
Figure 12: Out-of-plane angular correlations of coincident $\alpha$ particles
(circles) and protons (triangles) measured in the $^{28}$Si + $^{12}$C reaction
at E$_{lab}$ = 180 MeV. The proton correlations have been multiplied by a
factor 10$^{-2}$ for the sake of clarity. The ER's are detected at
$\Theta_{lab}$ = -10$^\circ$. The solid lines correspond to statistical-model
calculations discussed in the text. 
\end{figure}

\begin{figure}
Figure 13: Energy-correlation plots between coincident $\alpha$ particles and S
ER's produced in the $^{28}$Si + $^{12}$C reaction at E$_{lab}$ = 180 MeV. The
heavy fragment is detected at $\Theta_{S}$ = -10$^\circ$ and the
$\alpha$-particle angle settings are given in the figure. The dashed lines
correspond to different contours with their associated labellings discussed in
the text. 
\end{figure}

\begin{figure}
Figure 14: Experimental (left side) and calculated (right side)
energy-correlation plots between coincident $\alpha$ particles and S ER's
produced in the $^{28}$Si + $^{12}$C reaction at E$_{lab}$ = 180 MeV. The S is
identified at $\Theta_{S}$ = -10$^\circ$ and the $\alpha$ particles are
detected at $\Theta_{\alpha}$ = +40$^\circ$, and +70$^\circ$, respectively.
{\sc CACARIZO} calculations are discussed in the text. 
\end{figure}

\begin{figure}
Figure 15: Excitation-energy spectra calculated for the $^{28}$Si + $^{12}$C
reaction at E$_{lab}$ = 180 MeV for $^{8}$Be (left side) and $^{32}$S (right
side) in coincidence with the g.s. (above) and first excited level (below) of
$^{8}$Be. The solid line corresponds to the energy of the first excited state
of $^{8}$Be (3.08 MeV). The dashed lines correspond to an excitation energy in
$^{32}$S expected for an $\alpha$-transfer process. 
\end{figure}


\newpage

\begin{references}


\bibitem{Sanders99} S.~J. Sanders, A. Szanto de Toledo, and C. Beck, 
  \rm Phys. Rep. \bf 311\rm, 487 (1999). 

\bibitem{Matsuse97} T. Matsuse, C. Beck, R. Nouicer, and D. Mahboub, 
  \rm Phys. Rev. C {\bf 55}, 1380 (1997). 

\bibitem{Shivakumar87}B. Shivakumar, S. Ayik, B.~A. Harmon, and D. Shapira, 
  \rm Phys. Rev. C \bf35\rm, 1730 (1987). 

\bibitem{Shapira82}D. Shapira, R. Novotny, Y.~C. Chan, K.~A. Erb, J.~L.~C. Ford
jr., J.~C. Peng, and J.~D. Moses, \rm Phys. Lett. \bf 114B\rm, 111(1982); D.
Shapira, D. Schull, J.~L~.C. Ford, Jr., B. Shivakumar, R.~L. Parks, R.~A.
Cecil, and S.~T. Thornton, 
  \rm Phys. Rev. Lett. \bf 53\rm, 1634 (1984).

\bibitem{Shapira84} D. Shapira, D. Schull, J.~L.~C. Ford, Jr., B. Shivakumar,
R.~L. Parks, R.~A. Cecil and S.~T. Thornton,
  \rm Phys. Rev. Lett. \bf 53\rm, 1634 (1984).

\bibitem{Gary82} S. Gary and C. Volant,
  \rm Phys. Rev. C \bf 25\rm, 1877 (1982).

\bibitem{Lesko82} K.~T. Lesko, D.~-K. Lock, A. Lazzarini, R. Vandenbosch, V.
Metag, and H. Doubre,
  \rm Phys. Rev. C \bf 25\rm, 872 (1982).

\bibitem{Nagashima82} Y. Nagashima, S.~M. Lee, M. Sato, J. Schimizu, T.
Nakagawa, Y. Fukuchi, K. Furuno, M. Yamanouchi, and T.Mikumo,
  \rm Phys. Rev. C \bf 26\rm, 2661 (1982).

\bibitem{Harmon86} B.~A. Harmon, S.~T. Thornton, D. Shapira, J. Gomez del
Campo, and M. Beckerman, 
  \rm Phys. Rev. C \bf 34\rm, 552 (1986).

\bibitem{Harmon88} B.~A. Harmon, D. Shapira, P.~H. Stelson, B.~L. Burks, K.~A.
Erb, B. Shivakumar, K. Teh, and S.~T. Thornton, 
  \rm Phys. Rev. C \bf 38\rm, 572 (1988).

\bibitem{Vineyard93} M.~F. Vineyard {\it et al.},  
  \rm Phys. Rev. C {\bf 47}, 2374 (1993).

\bibitem{Arena94} N. Arena, Seb. Cavallaro, S. Femino, P. Figuera, S. Pirrone,
G. Politi, F. Porto, S. Romano, and S. Sambataro,
  \rm Phys. Rev. C \bf 50\rm, 880 (1994).

\bibitem{Pocanic85} D. Pocanic and N. Cindro,
  \rm Nucl. Phys. \bf A433\rm, 531 (1985); and references therein.

\bibitem{Ost79} R. Ost, M.~R. Clover, R.~M. DeVries, B.~R. Fulton, H.~E. Gove,
and N.~J. Rust,\\
  \rm Phys. Rev. C \bf 19\rm, 740 (1979).

\bibitem{Barrette79} J. Barrette, M.~J. LeVine, P. Braun-Munzinger, G.~M.
Berkowitz, M. Gai, J.~W. Harris, C.~M. Jachcinski, and C.~D. Uhlhorn,
  \rm Phys. Rev. C \bf 20,\rm 1759 (1979).

\bibitem{Beck94} C. Beck, Y. Abe, N. Aissaoui, B. Djerroud and F. Haas,
  \rm Phys. Rev. \bf C\rm 49, 2618(1994); and references therein.

\bibitem{Svensson00} C.~E. Svensson {\it et al.},   \rm Phys. Rev. Lett. \bf
85\rm, 2693 (2000);  \rm Nucl. Phys. \bf A682\rm, 1 (2001).

\bibitem{Svensson01} C.~E. Svensson {\it et al.},   \rm Phys. Rev. C \bf 63,\rm
 061301(R) (2001).

\bibitem{Ideguchi01} E. Ideguchi {\it et al.}, Phys. Rev. Lett. \bf 87,\rm 
222501 (2001).

\bibitem{Choudhury84} R.~K. Choudhury, P.~L. Gonthier, K. Hagel, M.~N.
Namboodiri, J.~B. Natowitz, L. Adler, S. Simon, S. Kniffen et G. Bergowitz, 
  \rm Phys. Lett. \bf B143\rm, 74 (1984).

\bibitem{Majka87} Z. Majka, M.~E. Brandan, D. Fabris, K. Hagel, A.
Menchaca-Rocha, J.~B. Natowitz, G. Nebbia, G. Prete, B. Sterling et G. Viesti,
  \rm Phys. Rev. C \bf35 \rm, 2125 (1987).

\bibitem{Govil87} I.~M. Govil, J.~R. Huizenga, W.~U. Schr\"oder, and J. T\"oke,
  \rm Phys. Lett. B \bf 197\rm, 515 (1987).

\bibitem{Fornal88} B. Fornal, G. Prete, G. Nebbia, F. Trotti, G. Viesti,
D. Fabris, K. Hagel, and J.~B. Natowitz,
  \rm Phys. Rev. C {\bf 37}, 2624 (1988).

\bibitem{Viesti88} G. Viesti, B. Fornal, D. Fabris, K. Hagel, J.~B. Natowitz,
G. Nebbia, G. Prete, and F. Trotti, 
  \rm Phys. Rev. C {\bf 38}, 2640 (1988).

\bibitem{Fornal89} B. Fornal, G. Viesti, G. Nebbia, and J.~B. Natowitz,
  \rm Phys. Rev. C {\bf 40}, 664 (1989).

\bibitem{Larana89} G. La Rana {\it et al.}, 
  \rm Phys. Rev. C {\bf 37}, 1920 (1988); {\it ibidem} C {\bf 40}, 2425 (1989).

\bibitem{Huizenga89} J.~R. Huizenga, A.~N. Behkami, I.~M. Govil, W.~U. 
Schr\"oder, and J. T\"oke,
  \rm Phys. Rev. C \bf 40\rm, 668 (1989).

\bibitem{Fornal91a} B. Fornal, F. Gramegna, G. Prete, R. Burch, G. D'Erasmo,
E.~M. Fiore, L. Fiore, A. Pantaleo, V. Paticchio, G. Viesti, P. Blasi, N. Gelli,
F. Lucarelli, M. Anghinolfi, P. Corvisiero, M. Taiuti, A. Zucchiatti, P.~F.
Bortignon, J. Ruiz, G. Nebbia, M. Gonin, and J.~B. Natowitz,\\
  \rm Phys. Lett. B {\bf 255}, 325 (1991).

\bibitem{Agnihotri93} D.~K. Agnihotri, A. Kumar, K.~C. Jain, K.~P. Singh, G.
Singh, D. Kabiraj, D.~K. Avasthi, and I.~M. Govil,
  \rm Phys. Lett. B \bf 307\rm, 283 (1993).

\bibitem{Govil98} I.~M. Govil, R. Singh, A. Kumar, J. Kaur, A.~K. Sinha, N.
Madhavan, D.~O. Kataria, P. Sugathan, S.~K. Kataria, K. Kumar, Bency John, and
G.~V. Ravi Prasad, 
  \rm Phys. Rev. C {\bf 57}, 1269 (1998).
 
\bibitem{Bandyopadhyay99} D. Bandyopadhyay, S.~K. Basu, C. Bhattacharya,
S. Bhattacharya, K. Krishan, A. Chatterjee, S Kailas, A. Navin, and
A. Shrivastava,
  \rm Phys. Rev. C \bf 59\rm, 1179 (1999).

\bibitem{Govil00a} I.~M. Govil, R. Singh, Ajay Kumar, G. Singh, S.~K. Kataria,
and S.~K. Datta,
  \rm Phys. Rev. C \bf 62\rm, 064606 (2000).

\bibitem{Govil00b} I.~M. Govil, R. Singh, A. Kumar, S.~K. Datta, and S.~K.
Kataria,
  \rm Nucl. Phys. \bf A674\rm, 377 (2000).

\bibitem{Bandy01} D. Bandyopadhyay, C. Bhattacharya, K. Krishan, S.
Bhattacharya, S.K. Basu, A. Chatterjee, S Kailas, A. Shrivastava, and  K.
Mahata,
  \rm Phys. Rev. C \bf 64\rm, 064613 (2001).

\bibitem{Fornal91b} B. Fornal {\it et al.},
  \rm Phys. Rev. C {\bf 44}, 2588 (1991).

\bibitem{Alamanos83} N. Alamanos, C. Levi, C. Le Metayer, W. Mittig, and L.
Papineau,
  \rm Phys. Lett. \bf 127B\rm, 23 (1983).

\bibitem{Bhattacharya99} C. Bhattacharya {\it et al.},  
  \rm Nucl. Phys. \bf A654\rm, 841c (1999).

\bibitem{Bhattacharya02} C. Bhattacharya, M. Rousseau, C. Beck, V. Rauch, R.M.
Freeman, D. Mahboub, R. Nouicer, P. Papka, O. Stezowski, A. Hachem, E. Martin,
A.K. Dummer, S.J. Sanders, and A. Szanto De Toledo,
  \rm Phys. Rev. C {\bf 65}, 014611 (2002). 

\bibitem{Rousseau00} M. Rousseau {\it et al.}, 
  \rm in Proceedings of the 7$^{th}$ International Conference on Clustering
Aspects of Nuclear Structure and Dynamics, eds. M. Korolija, Z. Basrak and R.
Caplar (World Scientific Publishing Co., Singapore, 2000), p.189. 

\bibitem{Beck00} C. Beck {\it et al.}, 
  \rm in Proceedings of the 9$^{th}$ International Conference on Nuclear
Reactions, Varenna, Italy, 2000, edited by E. Gadioli [Ricerca Scientifica ed
Educatione Permanente, Suppl. \bf 115\rm, (2000) p. 407]. 

\bibitem{Rousseau01a} M. Rousseau, 
  Ph.D.~thesis, Strasbourg University, Report {\bf IReS 01-02} (2001) \rm
(unpublished).

\bibitem{Rousseau01b} M. Rousseau {\it et al.}, 
  \rm in Proceedings of the XXXIX International Winter Meeting on Nuclear
Physics, Bormio, Italy, 2001, edited by I. Iori [Ricerca Scientifica ed
Educatione Permanente, Suppl. \bf 117\rm, (2001) p. 370.].

\bibitem{Bhattacharya01} C. Bhattacharya {\it et al.}, 

  \rm in Proceedings of the Workshop on Physics with Multidetector Arrays, Saha
Institute of Nuclear Physics, Calcutta, India, 2000, {\bf Pramana} Indian
Journal of Physics Vol. \bf 57\rm No. 1, (2001) p. 203.

\bibitem{Belier94} G. B\'elier,
  Ph.D.~thesis, Strasbourg University, Report {\bf CRN-94-34} (1994) \rm
\rm (unpublished). 

\bibitem{Bellot97} T. Bellot, 
  Ph.D.~thesis, Strasbourg University, Report {\bf IReS-97-35} (1997) \rm
(unpublithed). 

\bibitem{Kildir95} M. Kildir {\it et al.}, 
  \rm Phys. Rev. C {\bf 51}, 1873 (1995). 

\bibitem{Beck95} C. Beck {\it et al.}, Ricerca Scientifica ed Educazione
Permanente Supp. \bf 101\rm, 127 (1995).

\bibitem{Mahboub96} D. Mahboub,
  \rm Ph.D.~thesis, Strasbourg University, Report {\bf CRN-96-36} (1996) \rm
(unpublished).

\bibitem{Mahboub02} D. Mahboub {\it et al.}, to be published

\bibitem{Stokstad85} R.~G. Stokstad,
  \it Treatise on Heavy Ion Science \rm vol.3, ed. D.~A. Bromley (New York;
Plenum) (1985).

\bibitem{Cole00} A.~J. Cole,
  \it Statistical Models for Nuclear Decay,\rm ed. R.~R. Betts and W. Greiner
(IOP Publishing, Bristol and Philadelpia) (2000).

\bibitem{Charity00} R.~J. Charity,
  \rm Phys. Rev. C \bf 61\rm, 054614 (2000).

\bibitem{Puhlhofer77} F. P\"uhlhofer,
  \rm Nucl. Phys. \bf A280\rm, 267 (1977).

\bibitem{Wilmore64} D. Wilmore and P.~E. Hodgson,
  \rm Nucl. Phys. \bf 55\rm, 673 (1964).

\bibitem{Perey63} F.~G. Perey,
  \rm Phys. Rev. \bf 131\rm, 745 (1963); C.~M. Perey and F.~G. Perey,
  \rm Nucl. Data Tables \bf 17\rm, 1 (1976).

\bibitem{Huizenga61} J.~R. Huizenga,
  \rm Nucl. Phys. \bf 29\rm, 29 (1961).

\bibitem{Cohen74} S. Cohen, F. Plasil, and W.~J. Swiatecki,
  \rm Ann. Phys. (N.Y.) \bf 82\rm, 557 (1974).

\bibitem{Sierk86} A.~J. Sierk,
  \rm Phys. Rev. C \bf 33\rm, 2039 (1986).

\bibitem{Dilg73} W. Dilg, W. Schantl, H. Vonach, and M. Uhl,
Nucl. Phys. {\bf A217}, 269 (1973).

\bibitem{Shlomo91} S. Shlomo and J.~B. Natowitz, 
  \rm Phys. Rev. C {\bf 44}, 2878 (1991).

\bibitem{Toke81} J. T\"oke and W.~J. Swiatecki, 
  \rm Nucl. Phys. {\bf A372}, 141 (1981).

\bibitem{Janker99} P. J\"anker, H. Leitz, K.~E.~G. L\"obner, M. Morales, and
H.~G. Thies, 
  \rm Eur. Phys. J. A {\bf 4}, 147 (1999).

\bibitem{Myers66} W.D. Myers and W.J. Swiatecki,
  \rm Nucl. Phys. {\bf 81}, 1 (1966).

\bibitem{Viesti01} G. Viesti, V. Rizzi, D. Fabris, M. Lunardon, G. Nebbia, M.
Cinausero, b, E. Fioretto, G. Prete, A. Brondi, G. La Rana, R. Moro, E.
Vardaci, M. Aiche, M. M. Aleonard, G. Barreau, D. Boivin, J. N. Scheurer,
J. F. Chemin, K. Hagel, J. B. Natowitz, R. Wada, S. Courtin, F. Haas, N.
Rowley, B. M. Nyako, J. Gal and J. Molnar,
  \rm Phys. Lett. \bf 521B\rm, 165 (2001).

\bibitem{Blann81} M. Blann and T.~T. Komoto,
  \rm Phys. Rev. C {\bf 24}, 486 (1981).

\bibitem{Rudolph99} D. Rudolph {\it et al.},
  \rm Phys. Rev. Lett. \bf 82\rm, 3763 (1999);
  \rm Eur. Phys. J. A {\bf 4}, 115 (1999).

\bibitem{Ost80} R. Ost, A.J. Cole, M.R. Clover, B.R. Fulton and B.Sikora,
  \rm Nucl. Phys. \bf A342\rm, 185 (1980).

\bibitem{Gavron80} A. Gavron,
  \rm Phys. Rev. C \bf 21\rm, 230 (1980).

\bibitem{Mcdonald92} E.~W. MacDonald, A.~C. Shotter, D. Branford, J. Rahighi, 
T. Davinson, N.~J. Davis, Y. El-Mohri, and J. Yorkston,
  \rm Nucl. Instrum. Methods Phys. Res. \bf A317\rm, 498 (1992).

\bibitem{Ajzenberg88} F. Ajzenberg-Selove,
  \rm Nucl. Phys. \bf A490\rm, 1 (1988).

\bibitem{Arena95} N. Arena, Seb. Cavallaro, P. D'Agostino, G. Fazio, G.
Giardina, O.~Yu Goryunov, V.~V. Ostahko, R. Palamara, and A.~A. Shvedov,
  \rm J. Phys. G: Nucl. Part. Phys. \bf 21\rm, 1403 (1995).

\bibitem{Morgenstern86} H. Morgenstern, W. Bohne, W. Galster, and K. Grabisch,
  \rm Z. Phys. A \bf 324\rm, 443 (1986).

\bibitem{Nouicer99} R. Nouicer, C. Beck, R~.M. Freeman, F. Haas, N. Aissaoui,
T. Bellot, G. de France, D. Disdier, G. Duch\`ene, A. Elanique, A. Hachem,
F. Hoellinger, D. Mahboub, V. Rauch, S.~J. Sanders, A. Dummer, F.~W. Prosser,
A. Szanto de Toledo, Sl. Cavallaro, E. Uegaki, and Y. Abe,
  \rm Phys. Rev. C \bf 60\rm, 41303 (1999).

\bibitem{Beck01} C. Beck, R. Nouicer, D. Disdier, D. Duch\`ene, G. de France,
R.~M. Freeman, F. Haas, A. Hachem, D. Mahboub, V. Rauch, M. Rousseau, S.~J.
Sanders, and A. Szanto de Toledo,
  \rm Phys. Rev. C \bf 63\rm, 014607 (2001).

\bibitem{Thummerer00} S. Thummerer {\it et al.}, 
  \rm Phys. Scr. \bf T88\rm, 114 (2000).

\bibitem{Thummerer01} S. Thummerer, W. von Oertzen, B. Gebauer, D.~R. Napoli,
S.~M. Lenzi, A. Gadea, C. Beck, and M. Rousseau, 
  \rm J. Phys. G: Nucl. Part. Phys. \bf 27\rm, 1405 (2001).


\end{references}
\end{document}